\documentclass[aps,prc,twocolumn,groupedaddress]{revtex4-2}

\usepackage{amsmath}
\usepackage{graphicx}
\begin{document}


\title{Elastic scattering and total reaction cross sections of $^{6}$Li studied with a microscopic continuum discretized coupled channels model}


\author{Wendi Chen$^{1}$}
\author{D.Y. Pang$^{1}$}
\email[]{dypang@buaa.edu.cn}
\author{Hairui Guo$^{2}$}
\email[]{guo_hairui@iapcm.ac.cn}
\author{Tao Ye$^{2}$}
\author{Weili Sun$^{2}$}
\author{Yangjun Ying$^{2}$}
\affiliation{$^{1}$School of Physics, Beihang University, Beijing 100191, People’s Republic of China \\
$^{2}$ Institute of Applied Physics and Computational Mathematics, Beijing 100094, People’s Republic of China
}


\date{\today}

\begin{abstract}
We present a systematic study of $^{6}$Li elastic scattering and total reaction cross sections at incident energies around the Coulomb barrier within the continuum discretized coupled-channels (CDCC) framework, where $^{6}$Li is treated in an $\alpha$+$d$ two-body model. Collisions with $^{27}$Al, $^{64}$Zn, $^{138}$Ba and $^{208}$Pa are analyzed. The microscopic optical potentials (MOP) based on Skyrme nucleon-nucleon interaction for $\alpha$ and $d$ are adopted in CDCC calculations and satisfactory agreement with the experimental data is obtained without any adjustment on MOPs. For comparison, the $\alpha$ and $d$ global phenomenological optical potentials (GOP) are also used in CDCC analysis and a reduction no less than 50$\%$ on the surface imaginary part of deuteron GOP is required for describing the data. In all cases, the $^6$Li breakup effect is significant and provides repulsive correction to the folding model potential. The reduction on the surface imaginary part of GOP of deuteron reveals a strong suppression of the reaction probability of deuteron as a component of $^{6}$Li as compared with that of a free deuteron. A further investigation is made by taking the $d$ breakup process into account equivalently within the dynamic polarization potential approach and it shows that $d$ behaves like a tightly bound nucleus in $^{6}$Li induced reactions.
\end{abstract}


\maketitle

\section{introduction} \label{sec-1}

Understanding of nucleus-nucleus reaction mechanism is a permanent goal in nuclear physics. In particular, the nuclear reactions involving weakly bound projectiles have been a subject of great interest in recent years, as there are several unusual features compared with those induced by tightly bound nuclei \cite{Canto20061,Keeley2009396,Canto20151,Canto2020}. Due to the cluster structure with small separation energy in the weakly bound nucleus, its breakup probability is high and continuum states strongly influence other relevant observables such as elastic scattering angular distributions and total reaction cross sections.

Optical potential is a fundamental tool in the calculations of elastic scattering angular distributions and total reaction cross sections, which allows us to predict them in a simple way. In general, there are two kinds of optical potentials. One is the phenomenological optical potential, whose parameters should be adjusted to fit the experimental data. Another one is the microscopic optical potential (MOP), which is derived from nucleon–nucleon interaction theoretically. In the study of nuclear reactions induced by weakly bound nuclei, MOP plays an important role as the scattering measurement is not sufficient for obtaining a reliable global phenomenological optical potential (GOP) in many cases.

However, it is usually difficult to build a unified description with unadjusted MOP for the elastic scattering of weakly bound nuclei, as there are strong coupling effects from breakup, transfer and other reaction channels. For example, $^{6}$Li can break up into $\alpha$+$d$ and has a low threshold breakup energy of 1.47 MeV. Hence the breakup mechanisms are crucial in its induced reactions. There have been several measurements for $^{6}$Li elastic scattering, total reaction cross sections and other observables \cite{Figueira2007,Zadro2009,Maciel1999,Figueira2010} so far but the results are still far from being fully understood. The breakup threshold anomaly \cite{Hussein2006} has been observed in several experiments for $^6$Li projectile with a wide range of targets \cite{Canto20151}, in which the imaginary part of optical potential increases as the energy decreases to well below the Coulomb barrier. This anomaly is contributed by the breakup effect and results in a hindrance for MOP and GOP to well describe the elastic scattering angular distributions at energies around the Coulomb barrier.

Recently, an analysis \cite{Hamada2022} of $^6$Li+$^{90}$Zr elastic scattering shows that a reduction factor of $\sim$0.5 on the real part of MOP constructed with the S$\tilde{\mathrm{a}}$o Paulo potential \cite{Chamon2002,Chamon2021} is required to fit the data at energies below 2$V_B$, where $V_B$ is the height of the Coulomb barrier. This suppression is equivalent to adding a repulsive real polarization potential \cite{Hamada2022}, suggesting that the $^6$Li breakup effect should be taken into account for elastic scattering calculations.

So far, the continuum discretized coupled-channels method (CDCC) \cite{Austern1987125,Kamimura1986,Kawai1986a,Sakuragi1986a} has been one of the most popular methods to handle the breakup effect and investigate its role in the reactions induced by weakly bound nuclei. In this approach, a two-body model is generally adopted to describe the weakly bound nucleus, which is assumed to be composed of core and valence particles and will break up during the reaction. In early time, Sakuragi et al. \cite{Sakuragi1986a} presented a comprehensive CDCC study for $^6$Li elastic scattering at energies well above the Coulomb barrier with M3Y interaction \cite{Bertsch1977399}. They found that there is no need to adjust the real part of optical potential \cite{Bertsch1977399} in CDCC calculations, which is required to be reduced by $\sim$40$\%$ in the folding model potential analysis of Satchler and Love \cite{Satchler1979183}, however. As there have been quite a few measurements for $^6$Li scattering at energies around $V_B$ recently, it is worthwhile to provide a systematic study for $^6$Li induced reactions and investigate the effect of the breakup process.

Previously, we have obtained the nucleon-nucleus MOP based on Skyrme effective nucleon-nucleon interactions \cite{Shen2009} and constructed the MOPs for $^{2,3}$H, $^{3,4,6}$He and $^{7}$Li \cite{Guo2009c,Guo2010a,Guo2011d,Guo2014b,Chen2020,Chen2020a} with folding method. In this paper, we will combine our previous work of the MOP for light ion projectile with the CDCC method in the calculations of $^6$Li induced reactions at energies around $V_B$. This combination, named as microscopic CDCC, is not only an exploratory study to examining the validity of our MOP on the heavy ion projectile but also provides a possible way to apply our MOP on other heavier weakly bound projectiles, whose breakup effect is considerable, such as $^{11}$Li and $^{11}$Be.

On the other hand, $\alpha$ and $d$ phenomenological optical potentials have been widely adopted in the CDCC analysis for specific $^6$Li induced reactions. The adjustment on $d$ optical potential is generally accepted due to the effective suppression of the deuteron-target absorption \cite{Watanabe2012}. However, this adjustment is found to be dependent on the choice of $d$ optical potential. The surface part of $d$ optical potential is removed for $^{6}$Li+$^{209}$Bi reaction in Ref. \cite{Lei2015} but that is unchanged for $^{6}$Li+$^{159}$Tb and $^{6}$Li+$^{59}$Co reactions in Ref. \cite{Lei2017}. In the present work, we perform CDCC calculations with $\alpha$ and $d$ GOPs to ensure consistency on the adjustment of deuteron optical potential and compare their results with those calculated with MOPs to verify the validity of MOPs.

In addition, the 1$n$-stripping process has been found to be a significant contributor to the inclusive $\alpha$ cross sections in $^6$Li+$^{159}$Tb \cite{Pradhan2013} and $^{6}$Li+$^{112,124}$Sn \cite{Chattopadhyay2016,Parkar2023} reactions. In these cases, a neutron is stripped from $^6$Li and then $^5$Li is broken into $\alpha$+$p$. It would be interesting to investigate whether we should consider the breakup process of $d$ cluster in $^{6}$Li induced reactions as that in $(d,p)$ reactions \cite{Pang2013,Pang2014,Timofeyuk2020}. In this work, the breakup effect of $d$ is treated equivalently as a dynamic polarization potential. We put it into $^{6}$Li CDCC calculation to study the effect of $d$ breakup and check the validness of $\alpha$+$d$ two-body model for $^6$Li.

The paper is organized as follows. The theoretical framework is recapitulated in Sec. \ref{sec-2}. In Sec. \ref{sec-3}, we present the calculated results for $^6$Li-induced reactions and discuss the breakup effect of $^6$Li. The breakup possibility of $d$ cluster is discussed in Sec. \ref{sec-4}. A summary is given in Sec. \ref{sec-5}.

\section{Theoretical framework} \label{sec-2}

\subsection{CDCC formalism}\label{sec-2-1}

We recapitulate the three-body CDCC framework for $^6$Li scattering from a target nucleus (T). Details can be found in Refs. \cite{Austern1987125,Sakuragi1986a,Chen2022}. $\alpha$+$d$ two-body model is adopted to describe $^6$Li. Hence the total wave function of reaction system $\Psi$ with total energy $E$ is determined by the three-body Schr$\ddot{\mathrm{o}}$dinger equation
\begin{equation}\label{e-CDCC-eq}
  \left( H-E \right) \Psi =0,
\end{equation}
where $H$ is the total Hamiltonian,
\begin{equation}\label{e-H3b}
  H=T_R+U_{\alpha}+U_d+H_{\mathrm{in}}.
\end{equation}
$H_{\mathrm{in}}$ denotes the internal Hamiltonian of $^6$Li. $T_R$ represents the kinetic energy with respect to the relative coordinate $\boldsymbol{R}$ between $^6$Li and T. $U_x$ ($x$=$\alpha$, $d$) stands for the optical potential between $x$ and T.

In CDCC method, Eq. (\ref{e-CDCC-eq}) is solved in the model space spanned by the bound and discretized continuum states, which are all obtained by diagonalizing $H_{\mathrm{in}}$ with square-integrable basis functions. Therefore,
\begin{equation}\label{e-interH}
  H_{\mathrm{in}}\left| \psi _\gamma \right> =\varepsilon _\gamma\left| \psi _\gamma \right>
\end{equation}
where $\psi _\gamma$ is the $\gamma$-th eigenstate with eigenenergy $\varepsilon_\gamma$. For $^6$Li, $\psi _1$ represents the ground state and others denotes the discretized continuum states.

The total wave function $\Psi$ is expressed as
\begin{equation}\label{e-Psi}
  \Psi =\sum_{\gamma}{\chi _\gamma\left( \boldsymbol{R} \right) \left| \psi _\gamma \right>},
\end{equation}
where $\chi _\gamma$ represents the relative motion between T and $^6$Li in its $\gamma$-th state. $\chi _\gamma$ can be solved with the following coupled equations
\begin{equation}\label{e-CDCCeq-2}
  \left[ T_R+U_{\gamma \gamma}-\left( E-\varepsilon _\gamma \right) \right] \chi _\gamma=-\sum_{\gamma' \ne \gamma}^N{U_{\gamma \gamma'}\chi _{\gamma '}},
\end{equation}
where the coupling potential matrix is defined as
\begin{equation}\label{e-Unm}
  U_{\gamma \gamma'}=\left< \psi _\gamma \right|U_{\alpha}+U_d\left| \psi _{\gamma '} \right>.
\end{equation}

The Lagrange-mesh method is applied to diagonalize $H_{\mathrm{in}}$. Its basis functions are defined as
\begin{equation}\label{e-basis}
f_i\left( r \right) =\frac{\left( -1 \right) ^i}{\sqrt{hx_i}}\frac{L_N\left( r/h \right)}{r-hx_i}re^{-r/2h},i=1,2,...,N,
\end{equation}
where $\boldsymbol{r}$ is the relative coordinate between $\alpha$ and $d$. $L_N$ is the Laguerre polynomial of degree $N$. $x_i$ denotes the zeros of $L_N$, that is
\begin{equation}\label{e-zeros}
  L_N(x_i)=0, i=1,2,...,N.
\end{equation}
$h$ is a scaling parameter, adopted to the typical size of the system. See Refs. \cite{Druet201088,Baye20151,Chen2022} for details. The interaction between $\alpha$ and $d$ is taken from Refs. \cite{Chen2022,Chen2023}.


\subsection{$\alpha$ and $d$ optical potentials}\label{sec-2-2}

Microscopic optical potentials for $\alpha$ and $d$ clusters are constructed within folding model as
\begin{equation}\label{e-fold}
\begin{aligned}
U\left( \boldsymbol{R}_x \right) =\int{d\boldsymbol{s}} & \left[ U_n\left( \left| \boldsymbol{R}_x+\boldsymbol{s} \right| \right) \rho _{x,n}\left( \boldsymbol{s} \right) \right.
\\
& \left. + U_p\left( \left| \boldsymbol{R}_x+\boldsymbol{s} \right| \right) \rho _{x,p}\left( \boldsymbol{s} \right) \right],
\end{aligned}
\end{equation}
where $\boldsymbol{R}_x$ is the relative coordinate between $x$ and T. $\boldsymbol{s}$ is the internal coordinate of $x$. $\rho _{x,p}$ and $\rho _{x,p}$ denote the neutron and proton density distributions of $x$ respectively. $U_n$ and $U_p$ are isospin-dependent nucleon-nucleus MOPs \cite{Shen2009} based on Skyrme effective interaction. They are derived from the mass operator of the single-particle Green’s function through the nuclear matter approximation and the local density approximation. The real and imaginary parts of nucleon-nucleus MOP are denoted by the first- and the imaginary part of the second-order mass operators respectively. More details can be found in Ref. \cite{Shen2009}.

In this work, the Skyrme interaction SkMP \cite{Bennour1989} is used to calculate the nucleon-nucleus MOP. Improvements are made on the nucleon density distributions. The proton and neutron density distributions of targets are required for $U_n$ and $U_p$ calculations. Instead of Negele's formula \cite{Negele1970} in the original version of MOP \cite{Shen2009}, they are provided theoretically by an axially-symmetric self-consistent Dirac-Hartree-Bogoliubov mean field approach \cite{Carlson2000}, which has obtained good agreements with ground state masses and charge radii of near-stability nuclei throughout the entire mass table \cite{Chamon2021}. The proton and neutron density distributions of $\alpha$ are taken from Ref. \cite{Khoa2001} (version C in Table I) and those of $d$ are derived from its bound state wave functions, which is calculated with the $n$-$p$ Gaussian-type potential $V( r_{np} ) =-V_0\exp ( -r_{np}^{2}/r_{0}^{2} )$. $V_0$=72.25 MeV. $r_0$=1.484 fm. $r_{np}$ is the distance between $n$ and $p$.

For comparison, we also perform CDCC calculations with GOPs of $\alpha$ \cite{Avrigeanu2014} and $d$ \cite{An2006} and optical potential calculations with a $^6$Li GOP \cite{Xu2018}. The $\alpha$ GOP of Avrigeanu et al. \cite{Avrigeanu2014} and $d$ GOP of An et al. \cite{An2006} have been examined for their validities at low incident energies. They are suitable to be used in the present work.

\section{Results and discussion} \label{sec-3}

In the present work, the $^6$Li collisions with $^{27}$Al, $^{64}$Zn, $^{138}$Ba and $^{208}$Pb targets are analysed. The Coulomb barriers in the laboratory system are 8.0, 14.5, 22.0 and 30.6 MeV for these four reaction systems \cite{Figueira2007,Zadro2009,Maciel1999,Diaz-Torres2003} respectively. To investigate the breakup effect, we compare the full CDCC results with the one-channel (1ch) calculations, in which the continuum coupling is switched off.

\subsection{Calculation condition}

The first step for the analysis of $^6$Li induced reactions is to provide an adequate calculation condition for convergence. In the present work, we adopt the parameters $N$=35 and $h$=0.5 for the Lagrange-mesh method as in Ref. \cite{Chen2023}. Many tests have been performed to ensure that the calculated results are insensitive to $N$ and $h$.

In CDCC calculations, the angular momentum $l$ and energy $\varepsilon$ of $\alpha$-$d$ relative motion must be truncated at some $l_{\max}$ and $\varepsilon_{\max}$ values, which are chosen large enough so that the associated cross sections converge. As $l_{\max}$ and $\varepsilon_{\max}$ should be independent of the $\alpha$ and $d$ optical potentials, we are able to only examine the convergence of calculations with MOP and the final calculation condition can be applied to the calculations with GOP directly.

\begin{figure}[tbp]
\centering
\includegraphics[width=\columnwidth]{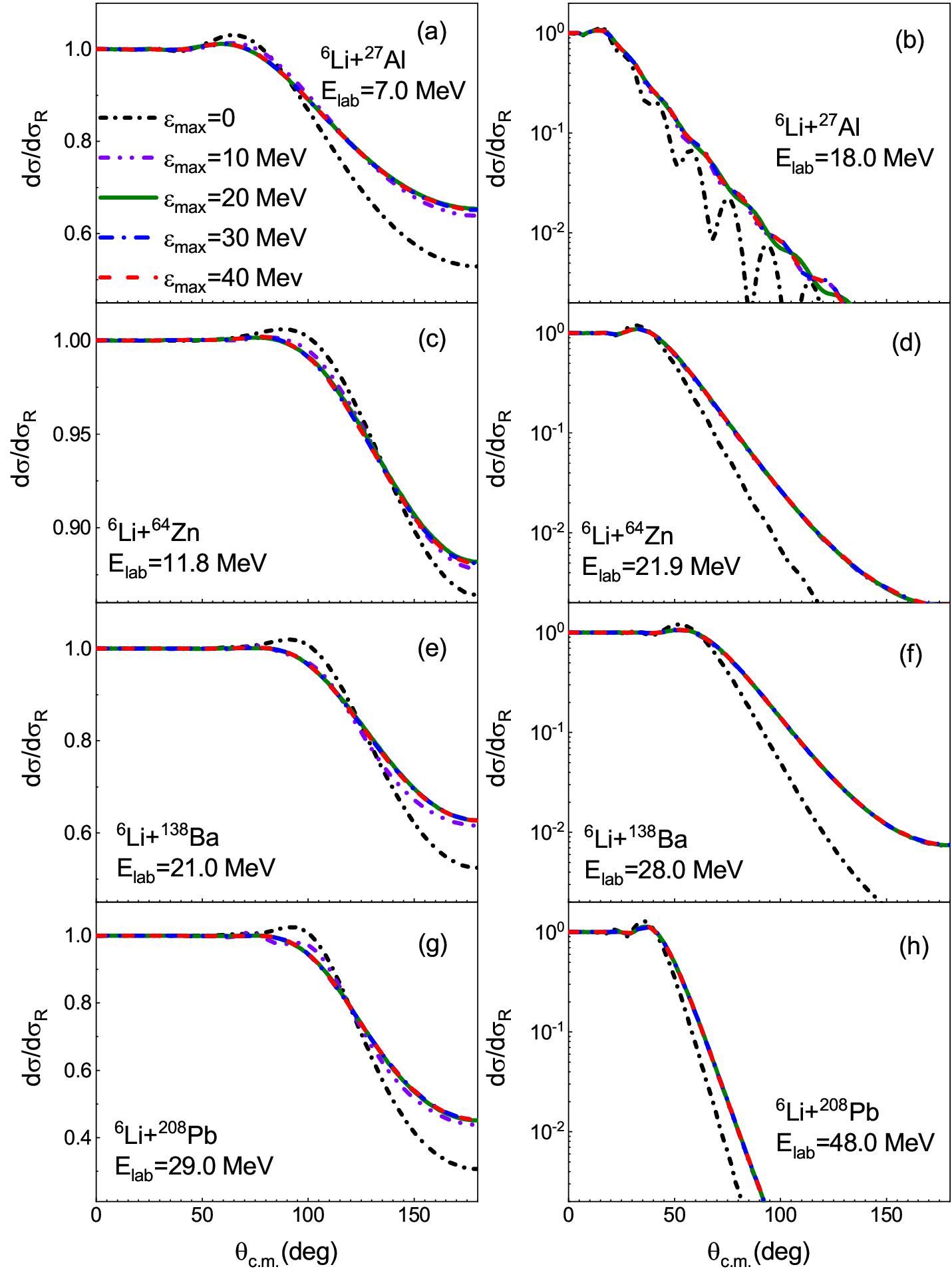}
\caption{Elastic scattering angular distributions, as ratios to Rutherford cross sections, for $^6$Li+$^{27}$Al, $^{64}$Zn, $^{138}$Ba and $^{208}$Pb systems at incident energies below and above the Coulomb barrier with different $\varepsilon_{\max}$. $l_{\max}$=2 is fixed. The Coulomb barriers in the laboratory system are 8.0, 14.5, 22.0 and 30.6 MeV for these four reaction systems \cite{Figueira2007,Zadro2009,Maciel1999,Diaz-Torres2003} respectively. The black, purple, green, blue and red lines denote the CDCC calculations with $\varepsilon_{\max}$=0, 10, 20, 30 and 40 MeV respectively. See text for details.}
\label{fig-elas-converge-emax}
\end{figure}

Firstly, we perform calculations for $^6$Li+$^{27}$Al, $^{64}$Zn, $^{138}$Ba and $^{208}$Pb systems at incident energies below and above the Coulomb barrier with different $\varepsilon_{\max}$. $l_{\max}$=2 is fixed so that the important $3^+$, $2^+$ and $1^+$ resonance states are included in CDCC calculations. Fig. \ref{fig-elas-converge-emax} shows the elastic scattering angular distributions calculated with $\varepsilon_{\max}$=0, 10, 20, 30 and 40 MeV. The CDCC calculation with $\varepsilon_{\max}$=0 is the one-channel calculation in fact. It can be observed for the $^{64}$Zn, $^{138}$Ba and $^{208}$Pb targets that the calculated results are almost the same with $\varepsilon_{\max} \geq$ 10 MeV at incident energies above the Coulomb barrier, while the convergence of elastic scattering angular distribution is reached with $\varepsilon_{\max} \geq$ 20 MeV in the sub-barrier energy region. However, the situation becomes more complicated for $^6$Li+$^{27}$Al reaction systems. The calculated results are the same with $\varepsilon_{\max} \geq$ 20 and 30 MeV at $E_{\mathrm{lab}}$=7.0 and 18.0 MeV respectively. A higher $\varepsilon_{\max}$ is required for the convergence of $^6$Li+$^{27}$Al elastic scattering angular distribution at incident energies above the Coulomb barrier.

\begin{figure}[tbp]
\centering
\includegraphics[width=\columnwidth]{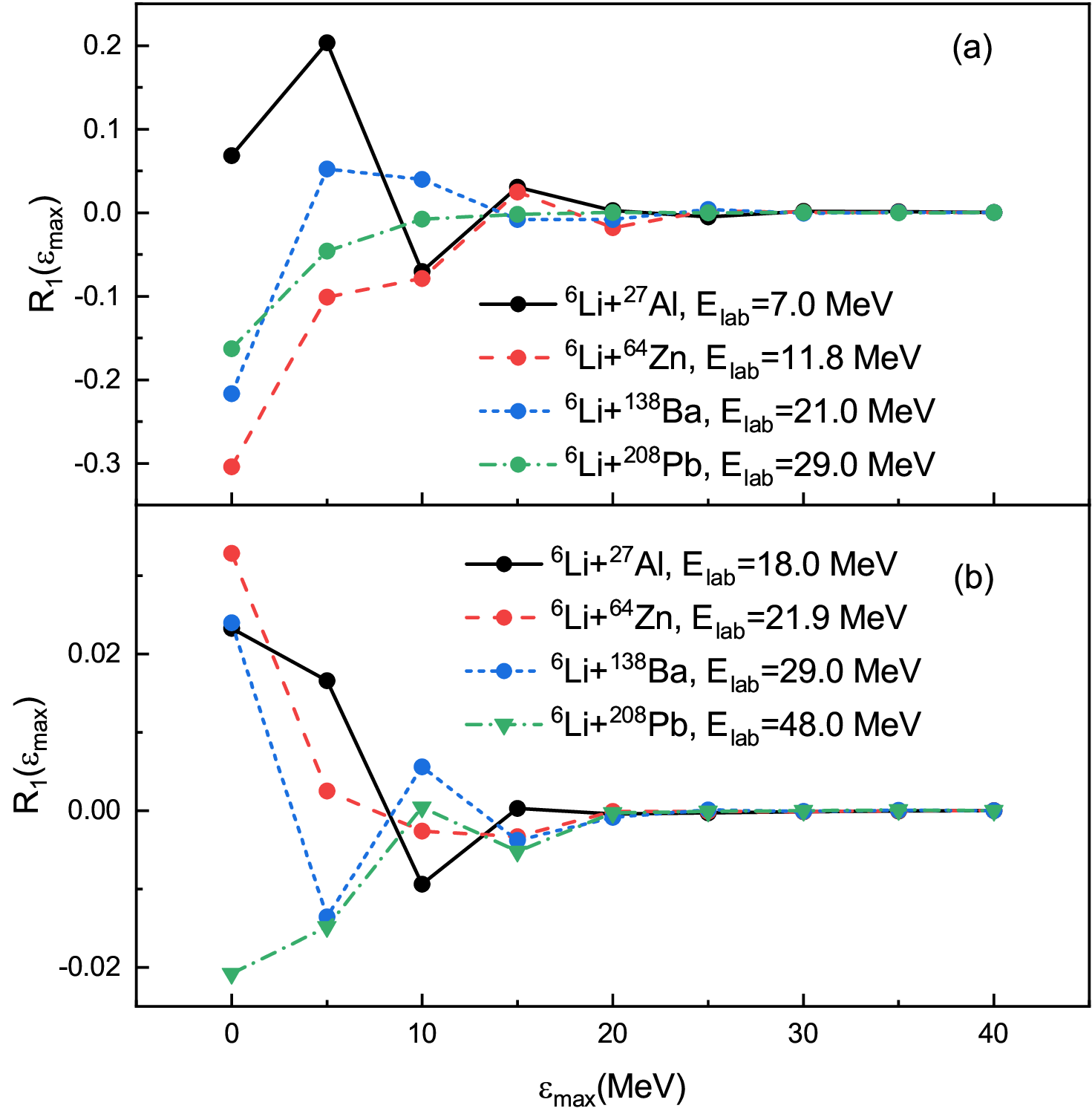}
\caption{$R_1(\varepsilon_{\max})$ for $^6$Li+$^{27}$Al, $^{64}$Zn, $^{138}$Ba and $^{208}$Pb systems shown with respect to $\varepsilon_{\max}$. $l_{\max}$=2 is fixed. (a) Incident energies are below the Coulomb barrier. (b) Incident energies are above the Coulomb barrier. The Coulomb barriers in the laboratory system are 8.0, 14.5, 22.0 and 30.6 MeV for these four reaction systems \cite{Figueira2007,Zadro2009,Maciel1999,Diaz-Torres2003} respectively. The solid, dashed, short dashed and dash-dotted lines denote the CDCC calculations for $^6$Li+$^{27}$Al, $^{64}$Zn, $^{138}$Ba and $^{208}$Pb systems respectively. See text for details.}
\label{fig-TR-converge-emax}
\end{figure}

We define $R_1(\varepsilon_{\max})$ as
\begin{equation}\label{e-R-emax}
  R_1(\varepsilon _{\max})=\frac{\sigma _{\mathrm{TR}} ( \varepsilon _{\max} )}{\sigma _{\mathrm{TR}} ( \varepsilon _{\max}=40 \, \mathrm{MeV} )}-1 ,
\end{equation}
where $\sigma _{\mathrm{TR}} (\varepsilon _{\max})$ is the total cross section calculated with $\varepsilon _{\max}$. $l_{\max}$=2 is hold. Fig. \ref{fig-TR-converge-emax} shows the value of $R_1(\varepsilon_{\max})$ as a function of $\varepsilon_{\max}$. For all reaction systems, the absolute values of $R_1(\varepsilon_{\max})$ are less than 2$\%$ when $\varepsilon_{\max} \geq$ 20 MeV. The convergence of total reaction cross section requires lower $\varepsilon_{\max}$ than that of elastic scattering angular distribution. With an overall consideration
, $\varepsilon_{\max}$=30 MeV is adopted to ensure the convergence of all CDCC calculations in the present work.

\begin{figure}[tbp]
\centering
\includegraphics[width=\columnwidth]{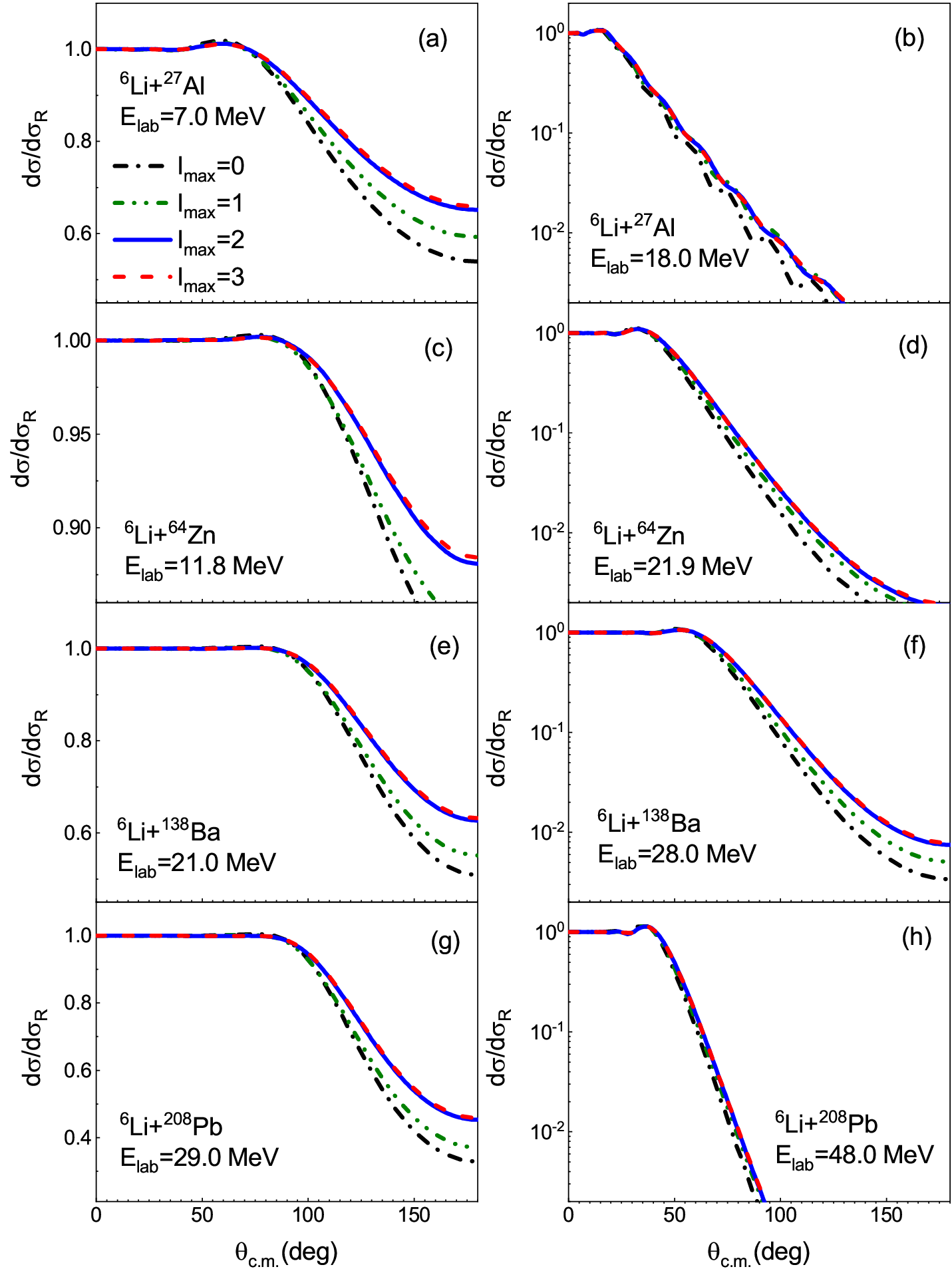}
\caption{Same as Fig. \ref{fig-elas-converge-emax} but the CDCC calculations are performed with different $l_{\max}$. $\varepsilon_{\max}$=30 MeV is fixed. The black, green, blue and red lines denote the CDCC calculations with $l_{\max}$=0, 1, 2 and 3 respectively. See text for details.}
\label{fig-elas-converge-lmax}
\end{figure}

Another important issue is the choice of $l_{\max}$. With fixed $\varepsilon_{\max}$=30 MeV, elastic scattering angular distributions are calculated with $l_{\max}$=0, 1, 2 and 3 for $^6$Li+$^{27}$Al, $^{64}$Zn, $^{138}$Ba and $^{208}$Pb systems at incident energies below and above the Coulomb barrier, shown in Fig. \ref{fig-elas-converge-lmax}. For all reaction systems, the elastic scattering angular distributions increase in the large scattering angle region as $l_{\max}$ increases from 0 to 2, while the differences between the calculated results with $l_{\max}$=2 and 3 are little.

\begin{figure}[tbp]
\centering
\includegraphics[width=\columnwidth]{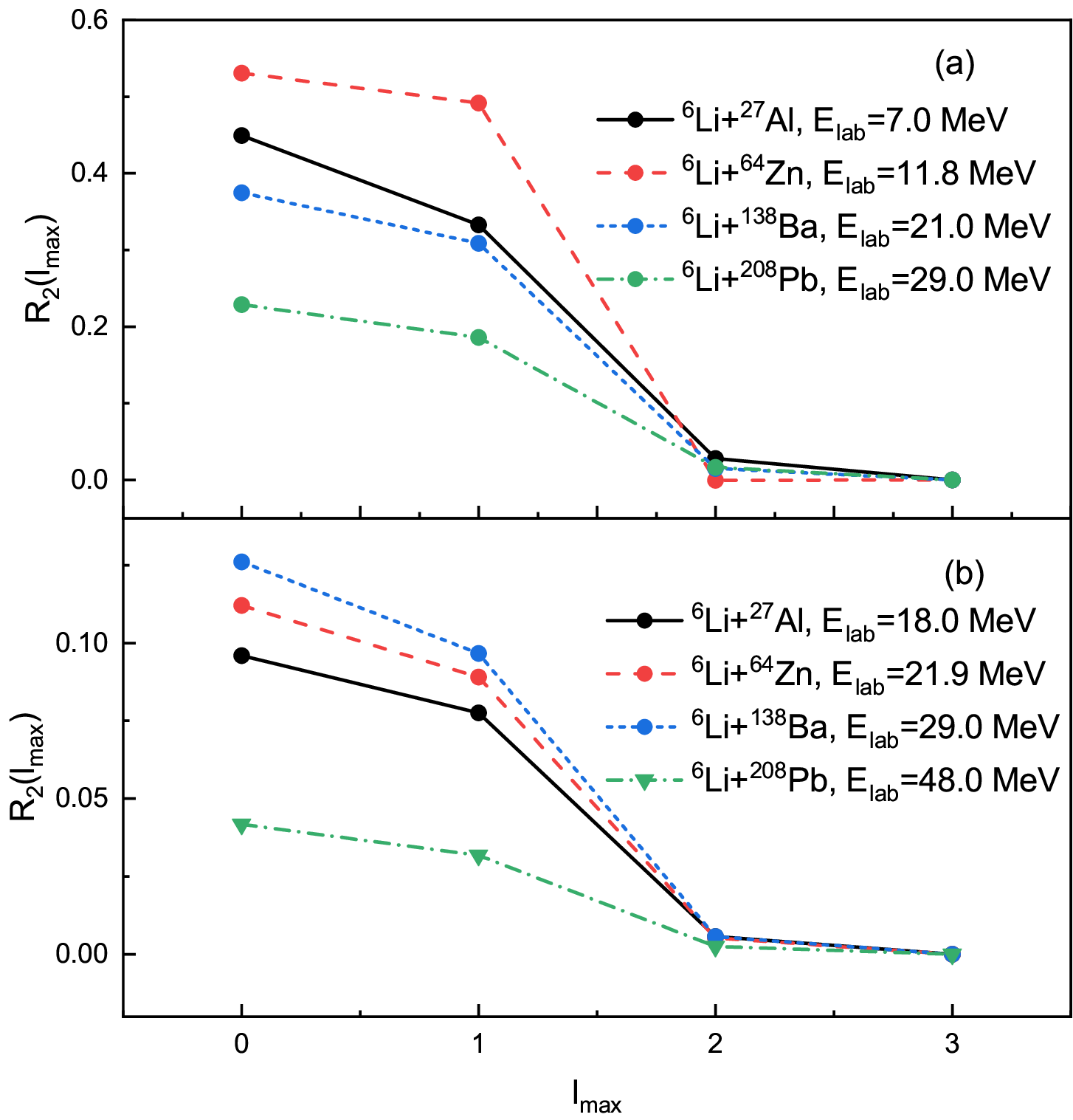}
\caption{Same as Fig. \ref{fig-TR-converge-emax} but for $R_2(l_{\max})$, shown as a function of $l_{\max}$. See text for details.}
\label{fig-TR-converge-lmax}
\end{figure}

Similar to $R_1(\varepsilon_{\max})$, we define $R_2(l_{\max})$ as
\begin{equation}\label{e-R-lmax}
  R_2(l _{\max})=\frac{\sigma _{\mathrm{TR}} ( l _{\max} )}{\sigma _{\mathrm{TR}} ( l _{\max}=3 )}-1 ,
\end{equation}
where $\sigma _{\mathrm{TR}} (l _{\max})$ is the total cross section calculated with $l _{\max}$. $\varepsilon_{\max}$=30 MeV is hold. $R_2(l_{\max})$ are presented in Fig. \ref{fig-TR-converge-lmax} as a function of $l_{\max}$. For all reaction systems, the value of $R_2(l_{\max})$ decreases as $l_{\max}$ increases and that becomes less than 2$\%$ at $l_{\max}$=2. Hence, $l_{\max}$=2 is enough to ensure the convergences of elastic scattering angular distribution and total reaction cross section.

In the following calculations, $\varepsilon_{\max}$=30 MeV and $l_{\max}$=2 are used.

\subsection{Reduction factor of $d$ GOP}\label{sec-3-1}

Generally, $^6$Li CDCC calculations with $\alpha$ and $d$ phenomenological optical potential require some renormalizations on the optical potentials as a consequence of the effective suppression of the deuteron-target absorption in $\alpha+d$ two-body model \cite{Watanabe2012,Watanabe2015}. Since the parameters of GOP are adjusted to fit various elastic scattering experimental data and total reaction cross sections, the CDCC analysis with GOP is expected to ensure the consistency of the renormalizations and improve reliability.

\begin{figure}[tbp]
\centering
\includegraphics[width=\columnwidth]{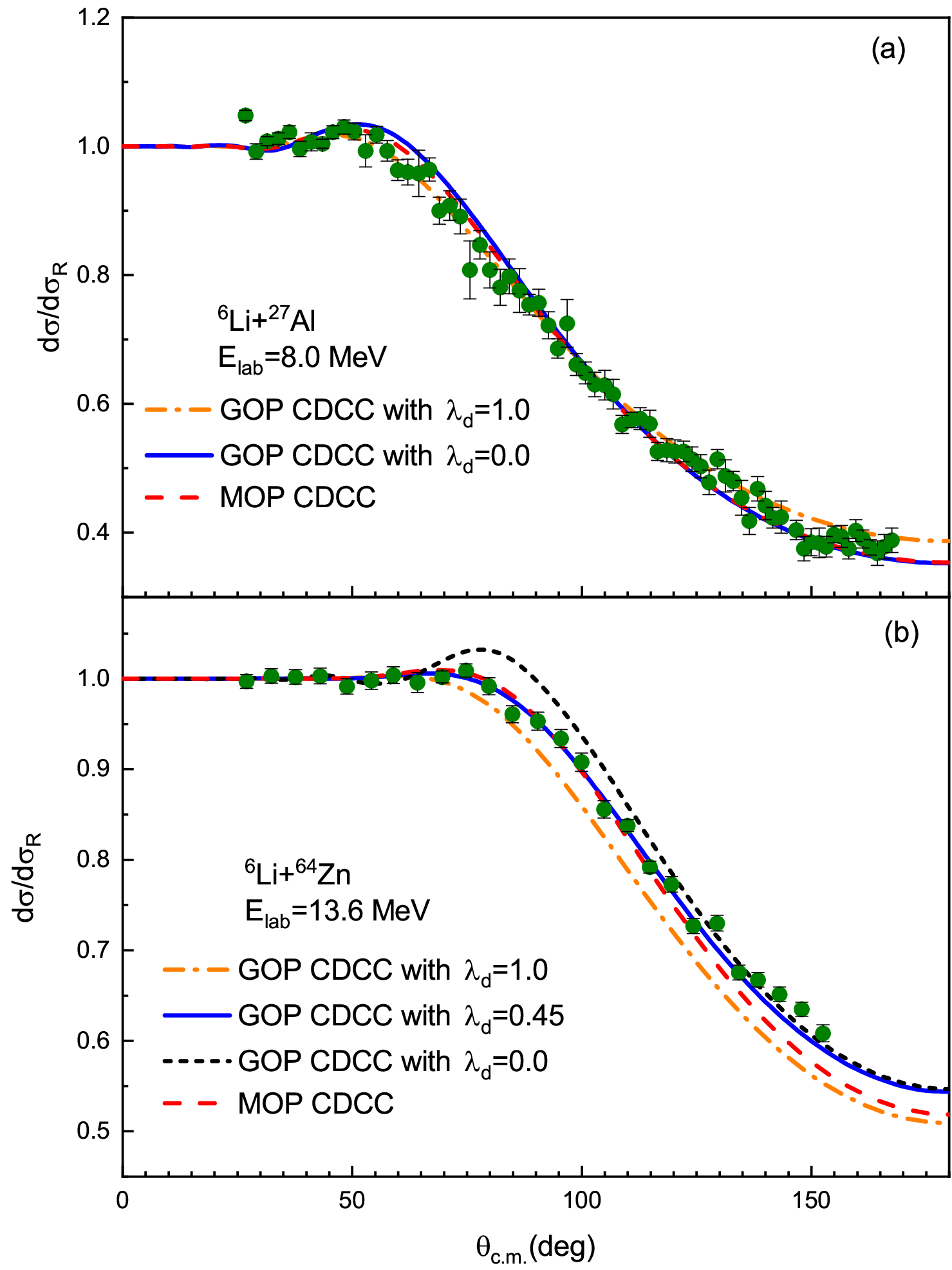}
\caption{Elastic scattering angular distributions, as ratios to Rutherford cross sections, for (a) $^6$Li+$^{27}$Al reaction at $E_{\mathrm{lab}}$=8.0 MeV and (b) $^6$Li+$^{64}$Zn reaction at $E_{\mathrm{lab}}$=13.6 MeV. The dash-dotted, solid and short dashed lines denote the CDCC calculations with GOP, in which the surface imaginary terms of $d$ GOP are multiplied by different $\lambda _d$ respectively. The dashed lines represent the CDCC results calculated with MOP. Experimental data are taken from Ref. \cite{Figueira2007,Zadro2009}. See text for details.}
\label{fig-factor-1}
\end{figure}

\begin{figure}[htbp]
\centering
\includegraphics[width=\columnwidth]{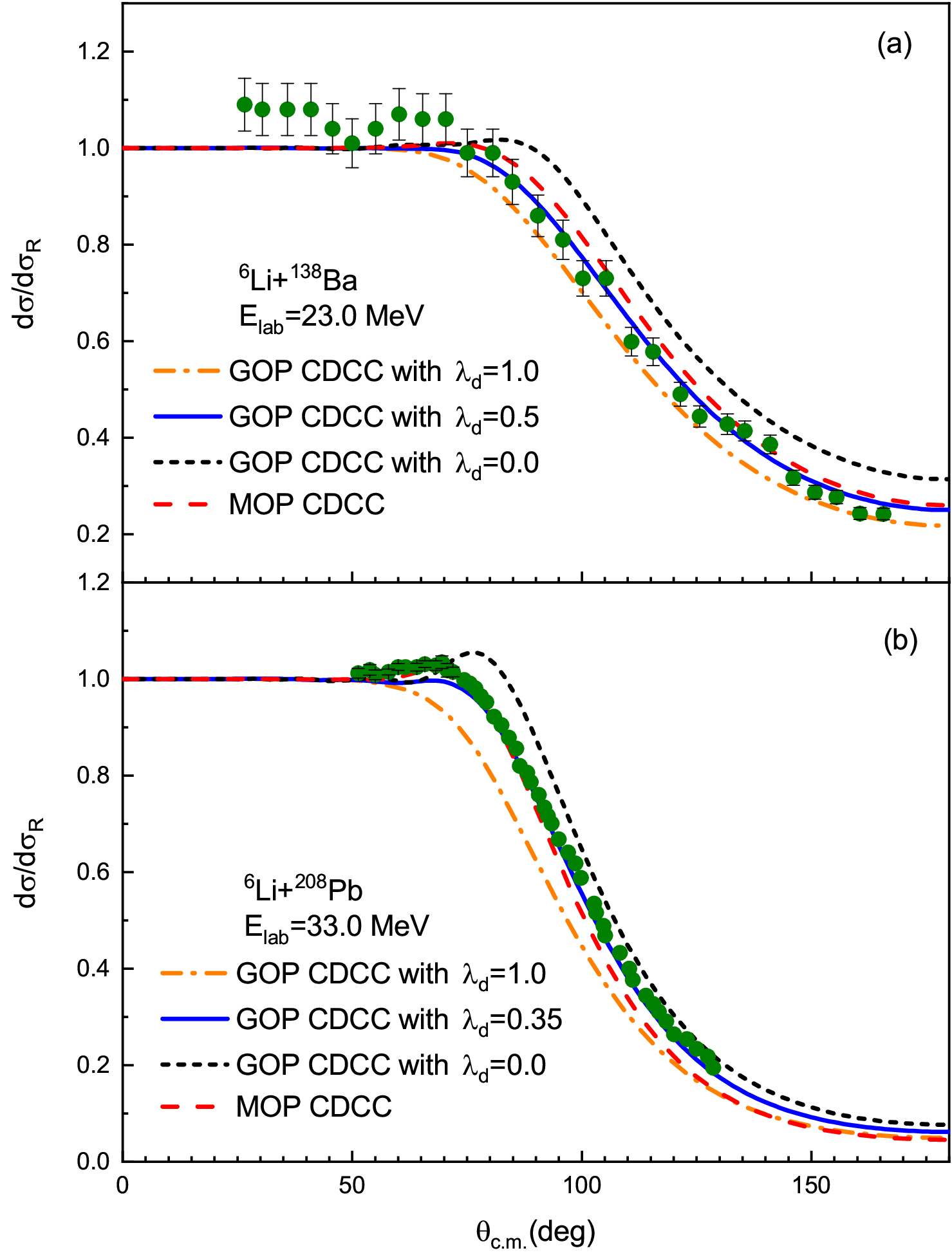}
\caption{Same as Fig. \ref{fig-factor-1} but for (a) $^6$Li+$^{138}$Ba reaction at $E_{\mathrm{lab}}$=23.0 MeV and (b) $^6$Li+$^{208}$Pb reaction at $E_{\mathrm{lab}}$=33.0 MeV. Experimental data are taken from Refs. \cite{Maciel1999,Keeley1994}. See text for details. }
\label{fig-factor-2}
\end{figure}

It is found that $^6$Li elastic scattering angular distribution is sensitive to the surface imaginary part of $d$ GOP, which should be multiplied by a reduction factor $\lambda _d$ to fit the experimental data. Considering the elastic scattering angular distributions and total cross sections simultaneously, $\lambda _d$ is optimized to be 0, 0.45, 0.50 and 0.35 for $^6$Li+$^{27}$Al, $^{64}$Zn, $^{138}$Ba and $^{208}$Pb reactions respectively. These reductions are held in the following calculations for $^6$Li induced reactions with $d$ GOP.

Figs. \ref{fig-factor-1} and \ref{fig-factor-2} show the CDCC calculations with GOP for $^6$Li+$^{27}$Al, $^{64}$Zn, $^{138}$Ba and $^{208}$Pb reactions at $E_{\mathrm{lab}}$=8.0, 13.6, 23.0 and 33.0 MeV respectively, compared with the results calculated with MOP. For $^6$Li+$^{64}$Zn, $^{138}$Ba and $^{208}$Pb reactions, with the decrease of $\lambda_d$ from 1 to 0, the deuteron-target absorption process is suppressed and the elastic scattering angular distribution increases in the angle region above 70 degrees, leading to a visible Coulomb rainbow \cite{Fricke1989399}. Good agreement between the experimental data \cite{Figueira2007,Zadro2009,Maciel1999,Keeley1994} and the CDCC results calculated with GOP is obtained with the optimized $\lambda _d$, which are no more than 0.5. However, for $^6$Li+$^{27}$Al reaction, the elastic scattering angular distribution changes moderately as $\lambda_d$ decreases from 1 to 0. We finally adopt $\lambda_d$=0 for this reaction system to fit the total reaction cross sections.

Different from the calculations with GOP, the CDCC calculations with MOP can reproduce the measured value well without any adjustment.

\subsection{$^6$Li elastic scattering}\label{sec-3-2}

\begin{figure}[htbp]
\centering
\includegraphics[width=\columnwidth]{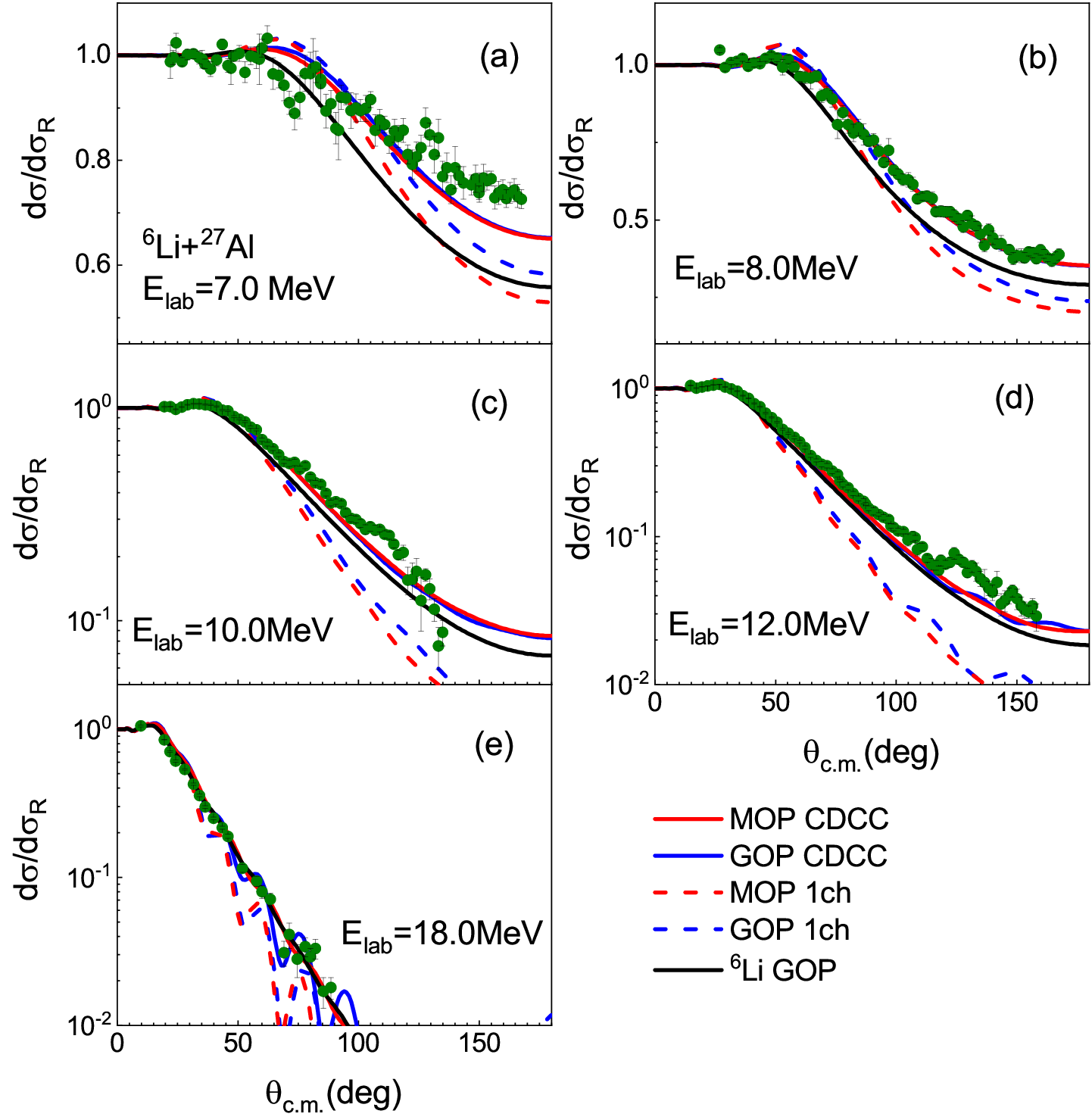}
\caption{Elastic scattering angular distributions, as ratios to Rutherford cross sections, for $^6$Li+$^{27}$Al reactions. The red solid and blue solid lines denote CDCC calculations with MOP and GOP respectively. The red dashed and blue dashed lines represent the one-channel calculations with MOP and GOP respectively. The black lines denote the results calculated with the $^6$Li GOP \cite{Xu2018}. The surface imaginary part of $d$ GOP is multiplied by $\lambda _d$=0 in the CDCC and one-channel calculations involving $d$ GOP. Experimental data are taken from Ref. \cite{Figueira2007} and represented by green circles. See text for details.}
\label{fig-6+27-elas}
\end{figure}

Fig. \ref{fig-6+27-elas} shows the calculated elastic scattering angular distributions for $^6$Li+$^{27}$Al reactions at incident energies ($E_{\mathrm{lab}}$) from 7.0 MeV to 18.0 MeV. Compared with one-channel results, continuum state coupling reduces the Coulomb rainbows and increases the angular distributions in the angle region above 70 degrees significantly at $E_{\mathrm{lab}}\leq$ 12.0 MeV. More specifically, two kinds of CDCC calculations provide almost the same results and obtain good agreements with the experimental data \cite{Figueira2007}, except the underestimations in back angle region at $E_{\mathrm{lab}}$=10.0 and 12.0 MeV. As the breakup reaction channel has been well included in CDCC calculations, these underestimations are considered to result from the lack of transfer reaction channel, which is another strong coupling channel in weakly bound nucleus induced reactions \cite{Pakou2022a}. At $E_{\mathrm{lab}}$=18.0 MeV, remarkable oscillations occur in the one-channel results while the CDCC results calculated with MOP are much smoother and agree with the measured value. In addition, the $^6$Li GOP \cite{Xu2018} underestimates the angular distributions above 50 degrees at $E_{\mathrm{lab}}\leq$ 10.0 MeV sizeably but provides similar results with CDCC method at higher incident energies.

\begin{figure}[htbp]
\centering
\includegraphics[width=\columnwidth]{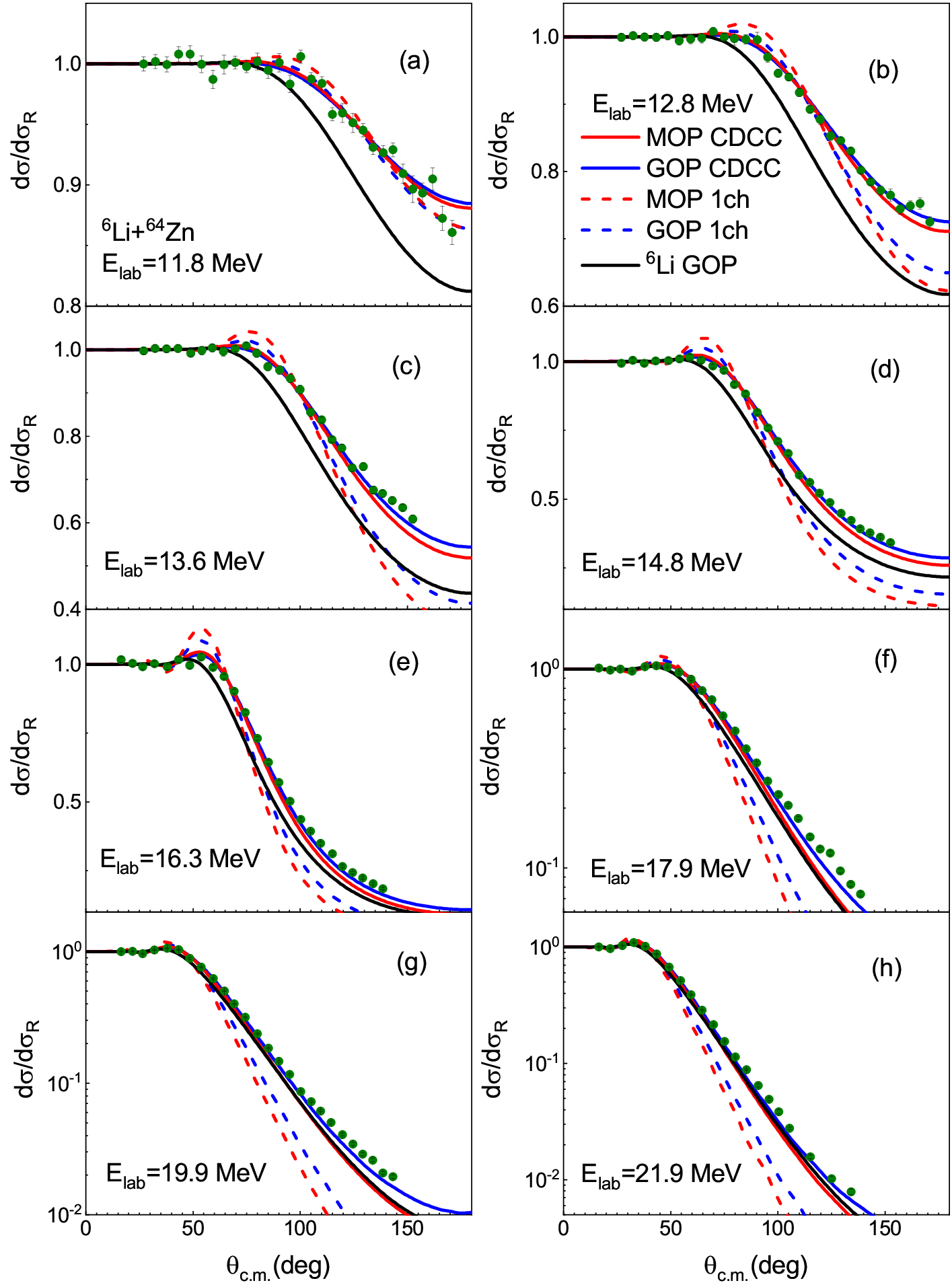}
\caption{Same as Fig. \ref{fig-6+27-elas} but for $^6$Li+$^{64}$Zn reactions. The surface imaginary part of $d$ GOP is multiplied by $\lambda _d$=0.45 in the CDCC and one-channel calculations involving $d$ GOP. Experimental data are taken from Ref. \cite{Zadro2009}.}
\label{fig-6+64-elas}
\end{figure}

The elastic scattering angular distributions for $^6$Li+$^{64}$Zn system at $E_{\mathrm{lab}}$=11.8-21.9 MeV are presented in Fig. \ref{fig-6+64-elas}. CDCC results with GOP are the same as those with MOP but become slightly larger in the back angle region, leading to a better agreement with experimental data \cite{Zadro2009} at $E_{\mathrm{lab}} \geq$ 13.6 MeV. Coulomb rainbows are obvious in one-channel calculations but they are negligible in CDCC results, which are strongly reduced by continuum state coupling.

\begin{figure}[htbp]
\centering
\includegraphics[width=\columnwidth]{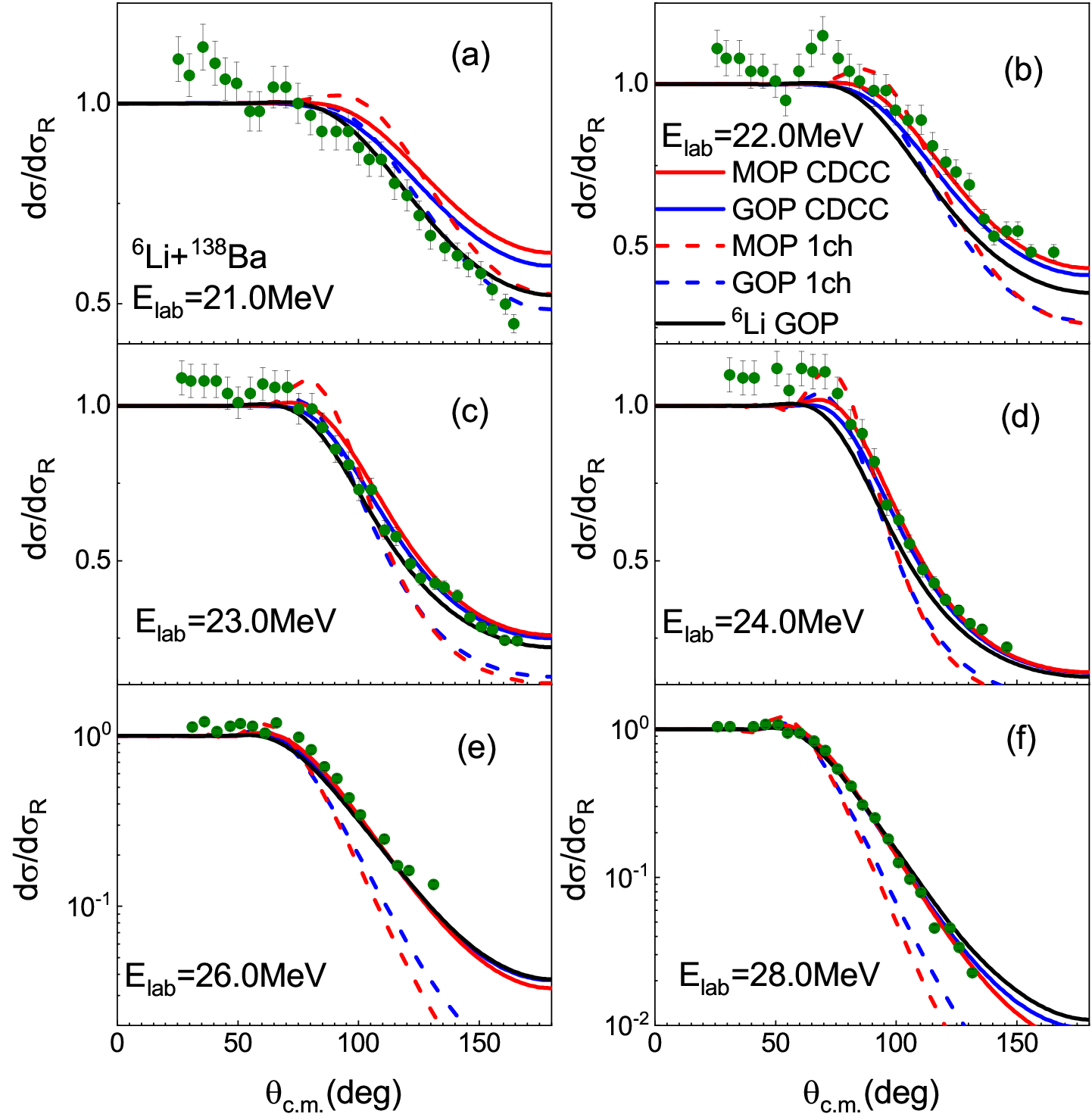}
\caption{Same as Fig. \ref{fig-6+27-elas} but for $^6$Li+$^{138}$Ba reactions. The surface imaginary part of $d$ GOP is multiplied by $\lambda _d$=0.5 in the CDCC and one-channel calculations involving $d$ GOP. Experimental data are taken from Ref. \cite{Maciel1999}.}
\label{fig-6+138-elas}
\end{figure}

\begin{figure}[htbp]
\centering
\includegraphics[width=\columnwidth]{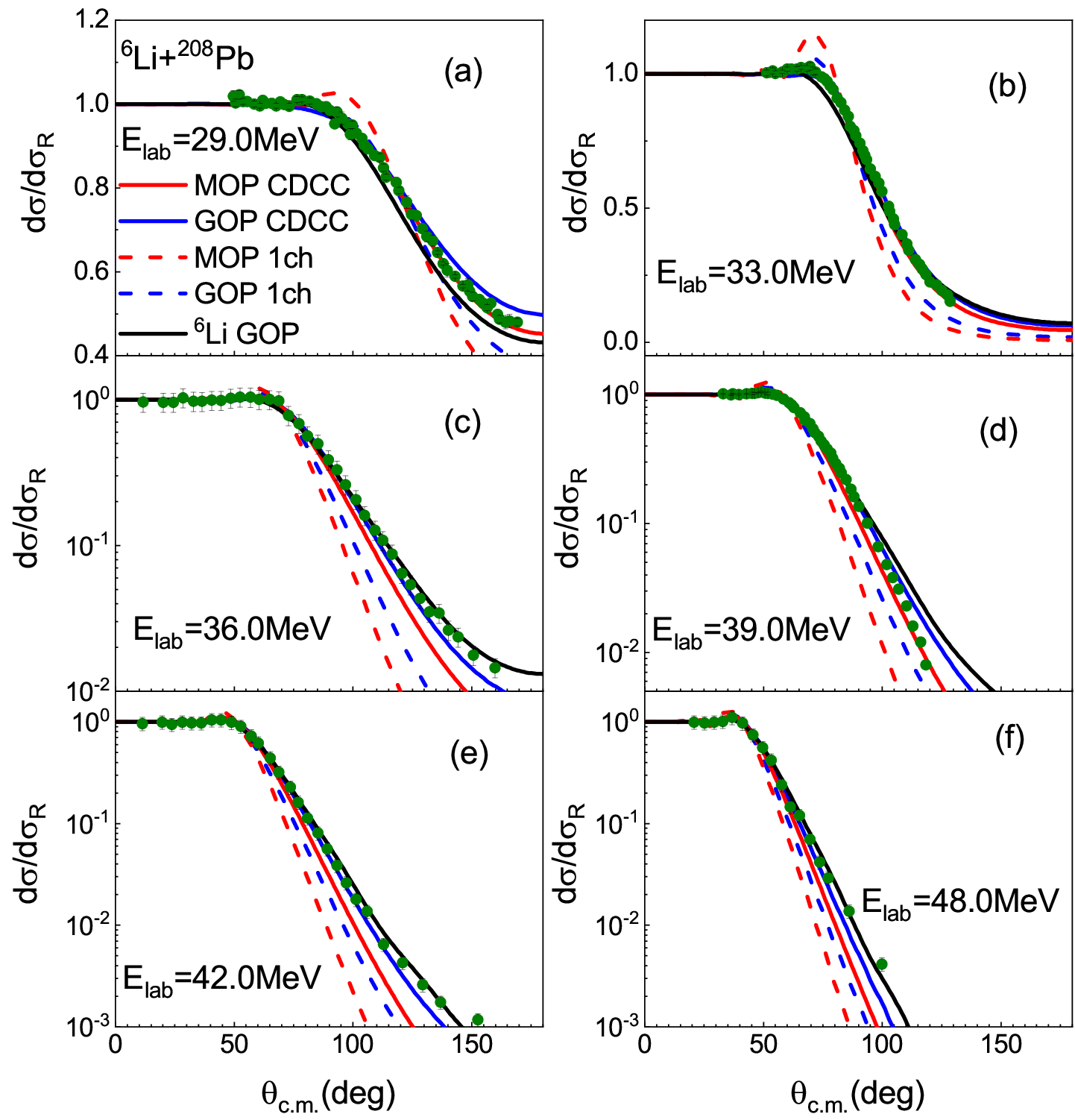}
\caption{Same as Fig. \ref{fig-6+27-elas} but for $^6$Li+$^{208}$Pb reactions. The surface imaginary part of $d$ GOP is multiplied by $\lambda _d$=0.35 in the CDCC and one-channel calculations involving $d$ GOP. Experimental data are taken from Refs. \cite{Gemmeke1978,Keeley1994}.}
\label{fig-6+208-elas}
\end{figure}

Figs. \ref{fig-6+138-elas} and \ref{fig-6+208-elas} show the $^6$Li elastic scattering on $^{138}$Ba and $^{208}$Pb targets. Consistent with the previous two reaction systems, continuum state coupling reduces the Coulomb rainbow and increases the angular distributions in the large angle region. CDCC results are in good agreement with experimental data \cite{Maciel1999,Gemmeke1978,Keeley1994} generally except for $^6$Li+$^{138}$Ba scattering at $E_{\mathrm{lab}}$=21.0 MeV. In addition, the CDCC calculations with GOP reproduce the measured value better than those with MOP for $^6$Li+$^{208}$Pb reactions as they provide larger angular distributions in the back angle region.

\begin{figure}[htbp]
\centering
\includegraphics[width=\columnwidth]{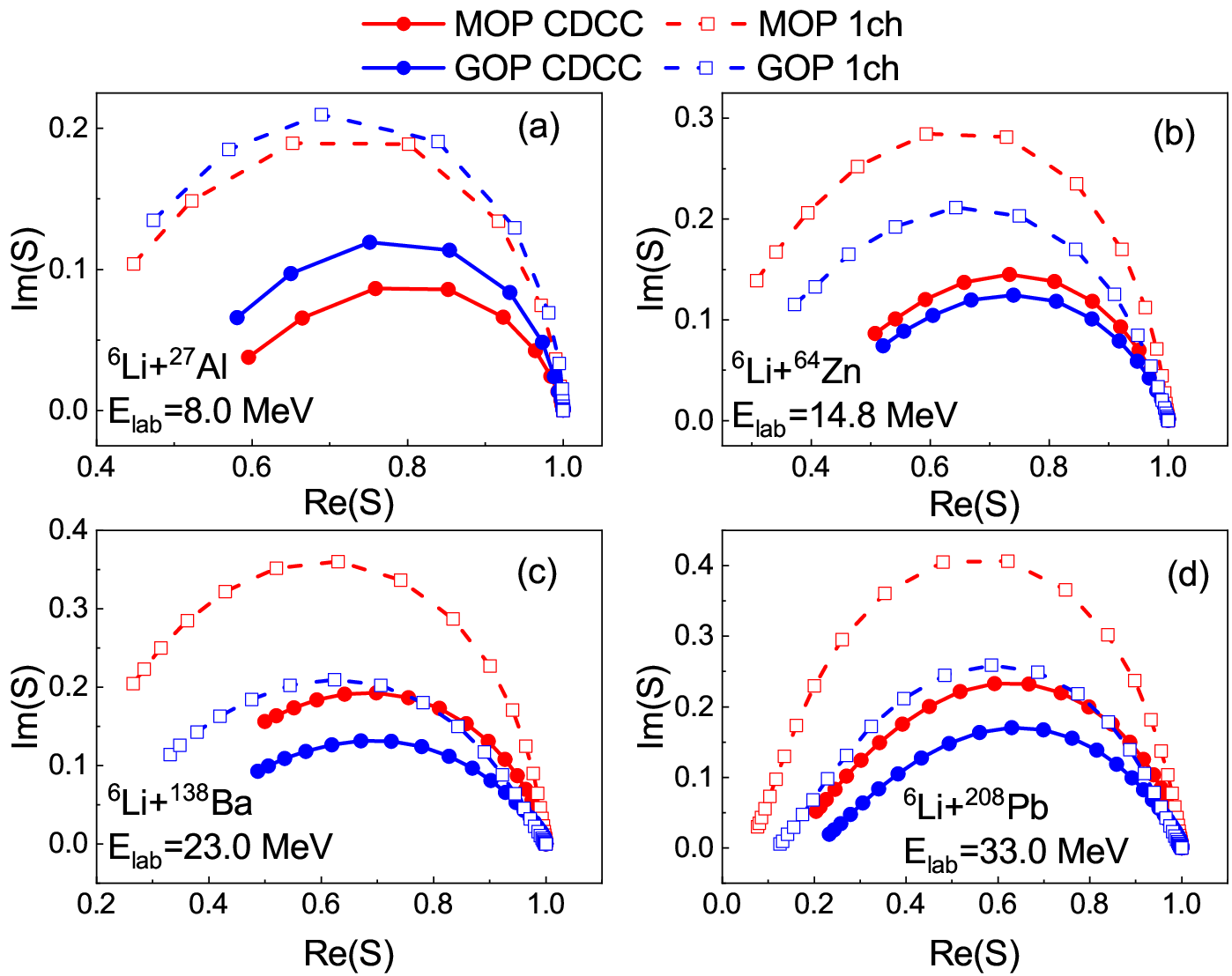}
\caption{Elastic scattering $S$-matrix elements $S^{J}_{JJ}$ for $^6$Li scattering on (a) $^{27}$Al at $E_{\mathrm{lab}}$=8.0 MeV, (b) $^{64}$Ni at $E_{\mathrm{lab}}$=14.8 MeV, (c) $^{138}$Ba at $E_{\mathrm{lab}}$=23.0 MeV and (d)$^{208}$Pb at $E_{\mathrm{lab}}$=33.0 MeV. Closed circles connected with solid lines and open squares connected with dashed lines stand for the results of CDCC and one-channel calculations respectively. Red and blue scatter lines represent the results obtained with MOP and GOP respectively. In the CDCC and one-channel calculations involving $d$ GOP, the surface imaginary parts of $d$ GOPs are multiplied by $\lambda _d$=0, 0.45, 0.5 and 0.35 for $^6$Li+$^{27}$Al, $^{64}$Zn, $^{138}$Ba and $^{208}$Pb systems respectively. See text for details.}
\label{fig-Smat}
\end{figure}

For a clearer understanding of the $^6$Li breakup effect on elastic scattering, we compare the elastic scattering $S$-matrix elements $S^{J}_{LL'}$ obtained from CDCC and one-channel calculations. $J$ is the total angular momentum. $L$ and $L'$ are the orbital angular momentums of initial and final channels with respect to $\boldsymbol{R}$. The diagonal elements $S^{J}_{JJ}$ are plotted in Fig. \ref{fig-Smat}. As $J$ increases, the module value of $S^{J}_{JJ}$ increases and $S^{J}_{JJ}$ converges to 1 finally. It can be observed at any $J$ that the real and imaginary parts of $S^{J}_{JJ}$ are enlarged and reduced by continuum state coupling respectively, which provides a repulsive correction to the results of one-channel calculations, that is, the folding potential. This correction is independent of optical potentials and reaction systems, consistent with that in the analysis of Sakuragi et al. \cite{Sakuragi1986a} at energies well above the Coulomb barrier.

\subsection{$^6$Li total reaction cross section}\label{sec-3-3}

\begin{figure}[htbp]
\centering
\includegraphics[width=\columnwidth]{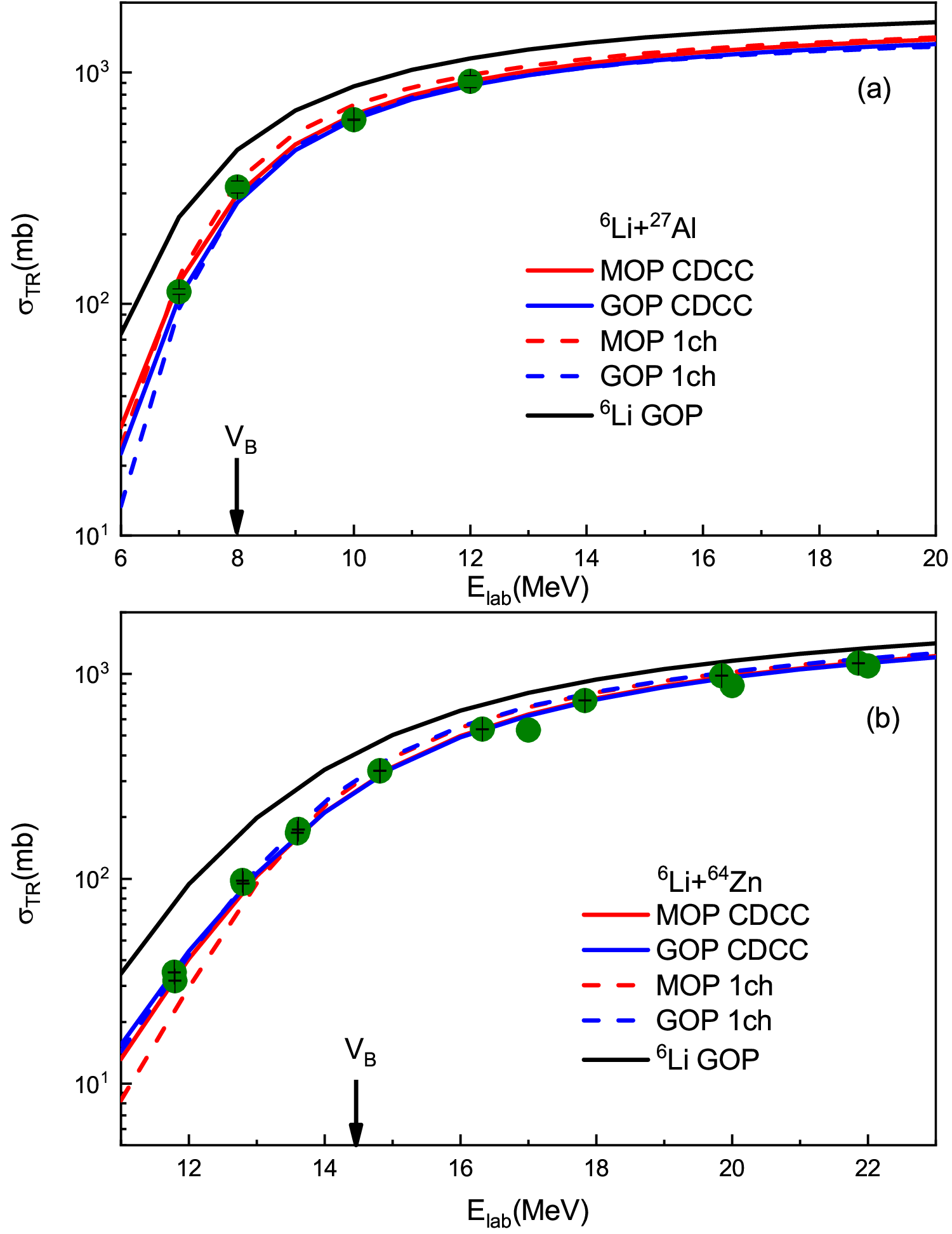}
\caption{Total reaction cross sections for (a) $^6$Li+$^{27}$Al and (b) $^6$Li+$^{64}$Zn systems. The red and blue solid lines denote CDCC calculations with MOP and GOP respectively. The red and blue dashed lines represent the one-channel calculations with MOP and GOP respectively. The black lines denote the results calculated with the $^6$Li GOP \cite{Xu2018}. In the CDCC and one-channel calculations involving $d$ GOP, the surface imaginary parts of $d$ GOPs are multiplied by $\lambda _d$=0 and 0.45 for $^6$Li+$^{27}$Al and $^{64}$Zn systems respectively. Experimental data are taken from Ref. \cite{Benjamim200730,Gomes2005,Zadro2009} and represented by green circles. The arrows indicate the Coulomb barriers in the laboratory system.}
\label{fig-total-cs-1}
\end{figure}

\begin{figure}[htbp]
\centering
\includegraphics[width=\columnwidth]{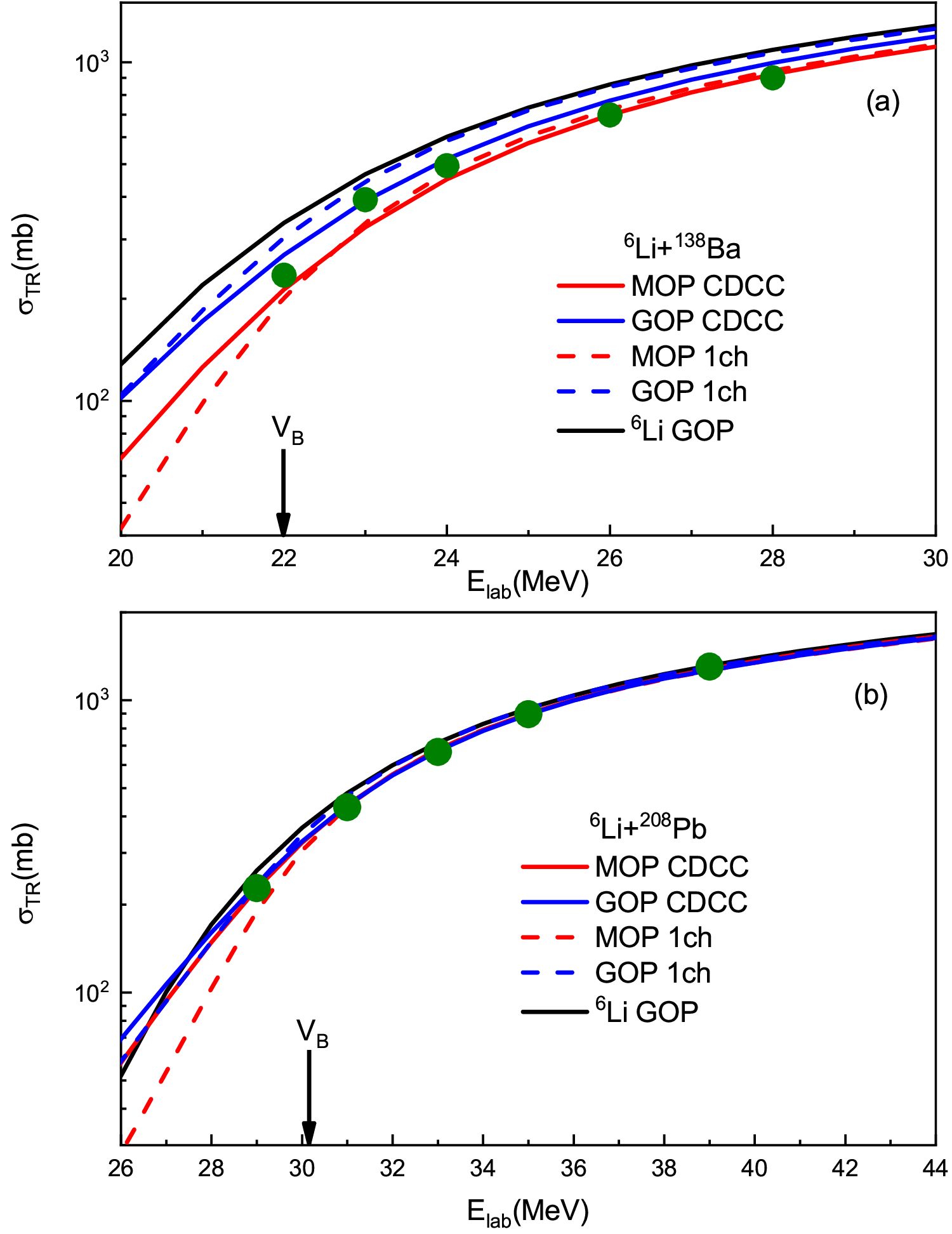}
\caption{Same as Fig. \ref{fig-total-cs-1} but for (a) $^6$Li+$^{138}$Ba and (b) $^6$Li+$^{208}$Pb systems.
In the CDCC and one-channel calculations involving $d$ GOP, the surface imaginary parts of $d$ GOPs are multiplied by $\lambda _d$=0.5 and 0.35 for $^6$Li+$^{138}$Ba and $^{208}$Pb systems respectively. Experimental data are taken from Refs. \cite{Maciel1999,Keeley1994}.}
\label{fig-total-cs-2}
\end{figure}

Comparison between the calculated total reaction cross sections and experimental data \cite{Benjamim200730,Zadro2009,Gomes2005,Maciel1999,Keeley1994} are presented in Figs. \ref{fig-total-cs-1} and \ref{fig-total-cs-2} for $^6$Li+$^{27}$Al, $^{64}$Zn, $^{138}$Ba and $^{208}$Pb systems. Although the data for $^6$Li+$^{27}$Al, $^{64}$Zn, $^{138}$Ba reaction systems are obtained by optical model potential analysis instead of experimental measurement, their values are reliable and can be adopted for comparison. CDCC results obtain satisfactory agreement with experimental data. In the calculations with MOP, continuum state coupling has a weak suppressive effect on $\sigma_{R}$ for all four reactions at energies above the Coulomb barrier but enlarges $\sigma_{\mathrm{TR}}$ considerably in the sub-barrier energy region for $^6$Li+$^{64}$Zn, $^{138}$Ba and $^{208}$Pb systems. These properties are held in the calculations with GOP but the enhancement on $\sigma_{\mathrm{TR}}$ at energies below the Coulomb barrier is moderate for $^6$Li+$^{27}$Al reaction and insignificant for $^6$Li scattering on heavier targets.

\section{Breakup process of $d$ cluster}\label{sec-4}

As presented in Sec. \ref{sec-3}, the surface imaginary part of $d$ GOP should be reduced no less than 50$\%$ to fit the data in the $^{6}$Li CDCC calculations with GOP, revealing a strong suppression on the $d$ reaction probability. It is well known that breakup and transfer reactions have strong coupling effects in $d$ induced reactions at low incident energies. Their contributions are roughly included in the surface imaginary part of $d$ phenomenological optical potentials, as these reactions mainly occur in the surface region. Hence it can be inferred that the $d$ cluster in $^6$Li is not able to break up as easily as that in the free state.

Since the 1$n$-stripping reaction contributes significantly to the inclusive $\alpha$ cross sections, we wish to investigate whether the breakup process of $d$ cluster should be taken into account in $^6$Li induced reactions. In the present work, $d$ breakup effect is treated approximately by means of dynamic polarization potential \cite{Sakuragi1986a,Cardenas2008}, which provides a correction to the folding model optical potential $U_d$=$U_d^{\mathrm{fold}}$.

Let us make an extreme hypothesis that $\alpha$ and $d$ clusters in $^{6}$Li just locate close to each other. They move with the same velocity at the beginning and react with the target simultaneously. By ignoring the excitation of target, the breakup process of $d$ can be described by the CDCC method at corresponding incident energy $E_d$=$\frac{1}{3}E_{\mathrm{lab}}$. For $^6$Li+$^{208}$Pb reaction at $E_{\mathrm{lab}}$=33.0 MeV, we perform CDCC calculation for $d$+$^{208}$Pb reaction at $E_d$=11.0 MeV with the microscopic nucleon optical potentials $U_n$ and $U_p$ as described in Sec. \ref{sec-2-2}. The $n$-$p$ interaction is the Gaussian-type potential as mentioned in Sec. \ref{sec-2-2}. Spins of nucleons are ignored. Following the method of Chau Huu-Tai \cite{Chauhuutai200656}, the $n$-$p$ continuum states of $S$, $P$ and $D$ partial waves in open channels are all taken into CDCC calculations. The Lagrange-mesh method is used to discretize continuum states with parameters $N$=30 and $h$=0.4 as suggested in Refs. \cite{Druet201088,Chen2022}.

With this CDCC calculation for $d$+$^{208}$Pb reaction, an $L$-independent polarization potential can be obtained with the method of Thompson et al. \cite{Thompson198984} as
\begin{equation}\label{e-Upol}
U_{d}^{\mathrm{pol}}=\frac{\sum_L{\left( 2L+1 \right) T_L}\left| \chi _L \right|^2U_{L}^{\mathrm{pol}}}{\sum_L{\left( 2L+1 \right) T_L}\left| \chi _L \right|^2},
\end{equation}
where $U_L^{\mathrm{pol}}$ is the trivially-equivalent $L$-dependent polarization potential calculated from the CDCC wave functions \cite{Franey1081,Sakuragi1986a}. $T_L$ and $\chi _L$ are the transmission coefficient and wave function of $d$-target relative motion in the elastic channel for the $L$-th partial wave respectively.

\begin{figure}[tbp]
\centering
\includegraphics[width=\columnwidth]{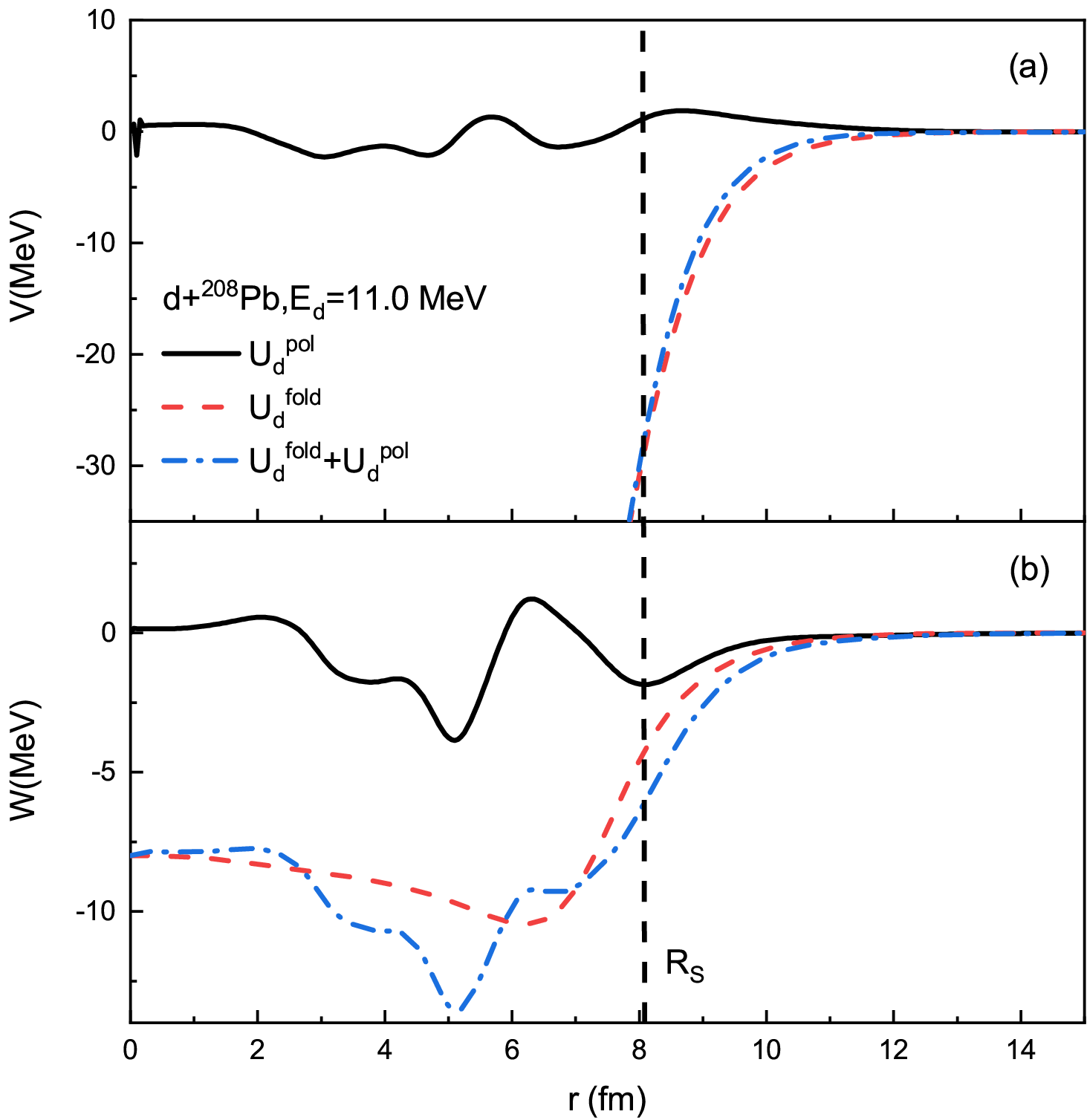}
\caption{Polarization potential $U_{d}^{\mathrm{pol}}$, folding model optical potential $U_{d}^{\mathrm{fold}}$ and their summation for $d$+$^{208}$Pb reaction at $E_d$=11.0 MeV. The solid, dashed and dash-dotted lines stand for $U_{d}^{\mathrm{pol}}$, $U_{d}^{\mathrm{fold}}$ and their summation respectively. The real and imaginary parts of potentials are presented in subfigures (a) and (b) respectively. The dashed vertical line indicates the interaction radius $R_S$=8.05 fm of the surface imaginary part of $d$ GOP \cite{An2006}.}
\label{fig-Upol}
\end{figure}

The polarization potential $U_{d}^{\mathrm{pol}}$, folding model potential $U_{d}^{\mathrm{fold}}$ and their summation are presented in Fig. \ref{fig-Upol}. In the low energy $d$ induced reactions, elastic scattering angular distribution is sensitive to the surface term of the optical potential but nearly independent of the inner part.  It can be seen that $U_{d}^{\mathrm{pol}}$ provides a repulsion correction to the real part of $U_d$ around $R_S$=8.05 fm, which is the interaction radius of the surface imaginary part of $d$ GOP \cite{An2006}. $U_{d}^{\mathrm{pol}}$ also deepens the imaginary part of $d$ optical potentials around $R_S$.

\begin{figure}[htbp]
\centering
\includegraphics[width=\columnwidth]{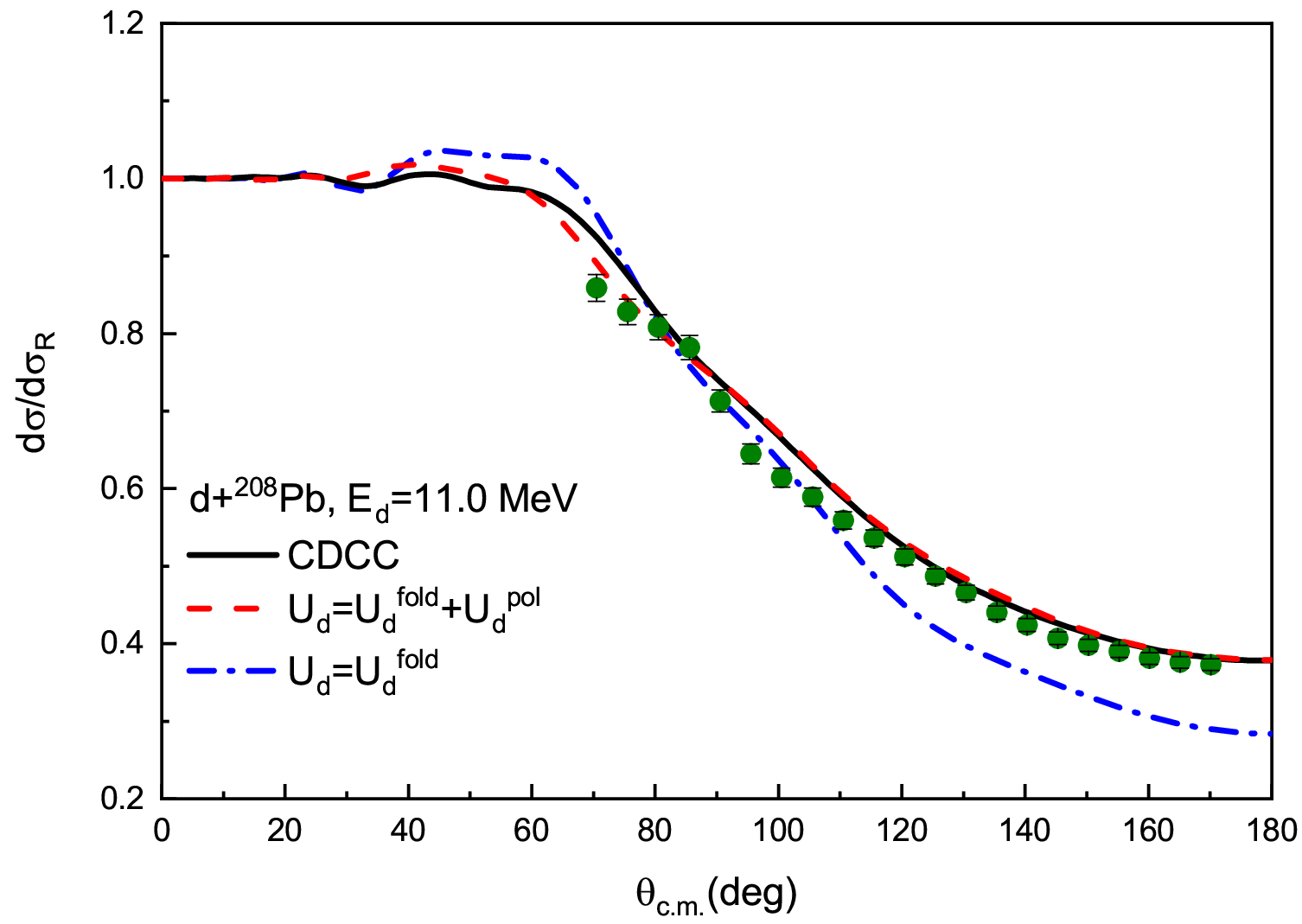}
\caption{Elastic scattering angular distributions, as ratios to Rutherford cross sections, for $d$+$^{208}$Pb reaction at $E_d$=11.0 MeV. The solid, dashed, dash-dotted lines represent the results of CDCC, $U_d^{\mathrm{fold}}$+$U_d^{\mathrm{pol}}$ and $U_d^{\mathrm{fold}}$ calculations. Experimental data are taken from Ref. \cite{Murayama1988261}. See text for details.}
\label{fig-d+208Pb-elas}
\end{figure}

Fig. \ref{fig-d+208Pb-elas} shows the elastic scattering angular distributions for $d$+$^{208}$Pb reaction at $E_d$=11.0 MeV. Compared with the results of $U_{d}^{\mathrm{fold}}$, continuum state coupling decreases the angular distributions in 40-80 degrees and increases the results above 80 degrees, improving the agreement with experimental data \cite{Murayama1988261}. Optical potential calculation with $U_{d}^{\mathrm{fold}}$+$U_{d}^{\mathrm{pol}}$ provides nearly the same results as those given by CDCC calculation. Hence $U_d^{\mathrm{pol}}$ is sufficient to hold the $d$ breakup effect when $d$ is in the free state.

\begin{figure}[htbp]
\centering
\includegraphics[width=\columnwidth]{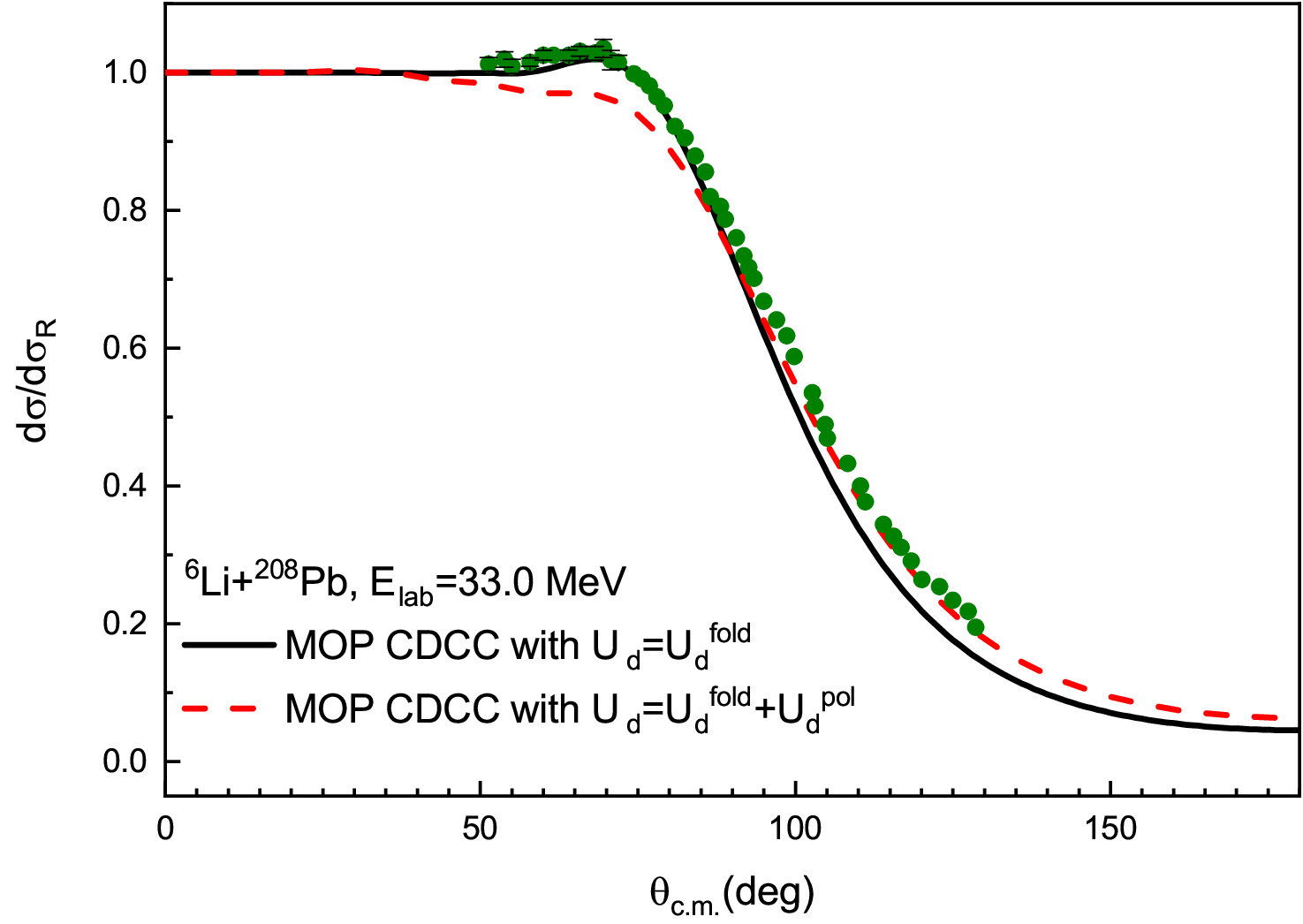}
\caption{Elastic scattering angular distributions, as ratios to Rutherford cross sections, for $^6$Li+$^{208}$Pb reaction at $E_{\mathrm{lab}}$=33.0 MeV. The solid and dashed lines represent CDCC results with deuteron optical potentials $U_{d}^{\mathrm{fold}}$ and $U_{d}^{\mathrm{fold}}$+$U_d^{\mathrm{pol}}$ respectively. Experimental data are taken from Ref. \cite{Keeley1994}. See text for details.}
\label{fig-6+208-33}
\end{figure}

Comparison between the CDCC results with deuteron optical potential $U_{d}^{\mathrm{fold}}$ and $U_{d}^{\mathrm{fold}}$+$U_d^{\mathrm{pol}}$ is presented in Fig. \ref{fig-6+208-33} for $^6$Li+$^{208}$Pb reaction at $E_d$=33.0 MeV. The $\alpha$ MOP remains unchanged. The angular distributions decrease in 40-90 degrees significantly and increase above 90 degrees slightly when the dynamic polarization potential $U_d^{\mathrm{pol}}$ is combined with folding model potential $U_{d}^{\mathrm{fold}}$ to be used as deuteron optical potential $U_d$. Although the agreement with experimental data \cite{Keeley1994} in the large scattering angle region is improved, this procedure badly underestimates the angular distributions in 40-90 degrees, suggesting that $U_d^{\mathrm{pol}}$ should not be added into $d$ optical potential. As the $U_{d}^{\mathrm{fold}}$ is constructed without the consideration of $d$ breakup process, it can be concluded that the $d$ cluster behaviours like a tight bound nucleus in $^6$Li induced reactions and there is no need to take the $d$ cluster breakup into account.

\section{Summary and conclusion} \label{sec-5}

We have systematically studied the $^6$Li elastic scattering and total reaction cross sections in the vicinity of the Coulomb barrier within the $\alpha$+$d$+target three-body framework for $^6$Li+$^{27}$Al, $^{64}$Zn, $^{138}$Ba and $^{208}$Pb reaction systems. A combination of the $\alpha$ and $d$ MOPs based on the Skyrme nucleon-nucleon effective interaction and CDCC method provides satisfactory agreement with the experimental data without any adjustment on optical potentials. In all cases, continuum state coupling suppresses the Coulomb barrier visibly and increases the elastic scattering angular distributions in the large scattering angle region, as it generates a repulsive correction to the folding model potential. Compared with the one-channel calculations in which the continuum state coupling is switched off, a slight suppression on the total cross section is observed in CDCC results for all four reaction systems at incident energies above the Coulomb barrier, while the total cross section in the sub-barrier energy region is enhanced considerably for $^6$Li+$^{64}$Zn, $^{138}$Ba and $^{208}$Pb systems.

For comparison, we also performed CDCC calculations with $\alpha$ and $d$ GOPs and optical potential calculations with $^6$Li GOP. The surface imaginary part of $d$ GOP should be reduced by no less than 50$\%$ to describe the data. CDCC results calculated with MOP and GOP are consistent with each other, suggesting that the MOP combined with the CDCC method is applicable for $^6$Li induced reactions.

The breakup possibility of the $d$ cluster has been investigated. A local dynamic polarization potential is generated to handle the breakup effect of $d$ and is added to the $d$ MOP to be the $d$ optical potential in $^6$Li CDCC calculation. This procedure significantly underestimates the angular distributions in the middle scattering angle region. It reveals that $d$ cluster behaviours like a tightly bound nucleus in $^6$Li induced reactions, consistent with the result of four-body CDCC analysis \cite{Watanabe2012,Watanabe2015} and resulting in the adjustment on the surface imaginary part of $d$ GOP. Hence, it would be practical to deal with the 1$n$-stripping reaction effectively in born approximation with the three-body model, instead of a full $\alpha$+$n$+$p$+target four-body computation. Related research is in progress.

\begin{acknowledgments}
 This work was supported by the National Natural Science Foundation of China unde Grant No. U2067205.
\end{acknowledgments}

\bibliography{apstemplate}

\end{document}